\documentclass[prx,aps,twocolumn,preprintnumbers,amsmath,amssymb,
superscriptaddress,showpacs]{revtex4-1}

\usepackage[colorlinks,bookmarks=true,citecolor=blue,linkcolor=red,urlcolor=blue,citecolor=blue]{hyperref}

\usepackage{subfigure}
\usepackage{color}

\usepackage{graphicx}
\usepackage{dcolumn}
\usepackage{bm}
\usepackage{braket}
\usepackage{fancyhdr,color}
\usepackage[table]{xcolor}
\usepackage{natbib}
\usepackage{tabularx}

\begin{document}

\title{Variational determination of arbitrarily many eigenpairs in one quantum circuit}

\author{Guanglei Xu}\email{These authors contributed equally to this work}
\affiliation{Institute of Physics,
Chinese Academy of Sciences, Beijing 100190, China.}

\author{Yi-Bin Guo}\email{These authors contributed equally to this work}

\author{Xuan Li}
\affiliation{Institute of Physics,
Chinese Academy of Sciences, Beijing 100190, China.}
\affiliation{School of Physical Sciences, University of Chinese Academy of Sciences, Beijing 100049, China.}

\author{Zong-Sheng Zhou}
\affiliation{Institute of Physics,
Chinese Academy of Sciences, Beijing 100190, China.}

\author{Hai-Jun Liao}\email{navyphysics@iphy.ac.cn}
\affiliation{Institute of Physics,
Chinese Academy of Sciences, Beijing 100190, China.}
\affiliation{Songshan Lake Materials Laboratory, Dongguan, Guangdong 523808, China.}

\author{T. Xiang}\email{txiang@iphy.ac.cn}
\affiliation{Institute of Physics, Chinese Academy of Sciences, Beijing 100190, China.}
\affiliation{Beijing Academy of Quantum Information Sciences, Beijing, China.}
\affiliation{School of Physical Sciences, University of Chinese Academy of Sciences, Beijing 100049, China.}

\begin{abstract}
The state-of-the-art quantum computing hardware has entered the noisy intermediate-scale quantum (NISQ) era. Having been constrained by the limited number of qubits and shallow circuit depth, NISQ devices have nevertheless demonstrated the potential of applications on various subjects. One example is the variational quantum eigensolver (VQE) that was first introduced for computing ground states. Although VQE has now been extended to the study of excited states, the algorithms previously proposed involve a recursive optimization scheme which requires many extra operations with significantly deeper quantum circuits to ensure the orthogonality of different trial states. Here we propose a new algorithm to determine many low energy eigenstates simultaneously. By introducing ancillary qubits to purify the trial states so that they keep orthogonal to each other throughout the whole optimization process, our algorithm allows these states to be efficiently computed in one quantum circuit.  Our algorithm reduces significantly the complexity of circuits and the readout errors, and enables flexible post-processing on the eigen-subspace from which the eigenpairs can be accurately determined. We demonstrate this algorithm by applying it to the transverse Ising model. By comparing the results obtained using this variational algorithm with the exact ones, we find that the eigenvalues of the Hamiltonian converge quickly with the increase of the circuit depth. The accuracies of the converged eigenvalues are of the same order, which implies that the difference between any two eigenvalues can be more accurately determined than the eigenvalues themselves.

\end{abstract}

\maketitle

\section{Introduction}
Quantum computation and quantum simulation, which utilize quantum devices to solve classical or quantum problems, hold promises to tackle large scale systems which are intractable with conventional computers. Several quantum algorithms have been proved to have exponential or quadratic  speed-up over classical counterparts~\cite{shor1994,lloyd1996,harrow2009,grover1996}. This stimulated the exploration of quantum hardwares~\cite{divincenzo1995,divincenzo2000}.
The implementation of promising quantum algorithms requires enormous number of reliable qubits and quantum operations with high fidelity, many platforms such as superconducting circuits~\cite{blais2021}, ultracold atoms~\cite{bloch2008}, trapped ion systems~\cite{monroe2021}, photonic systems~\cite{kok2007}, and nuclear magnetic resonance with nitrogen-vacancy spin systems~\cite{mamin2013} have demonstrated the potential in quantum computing.

Quantum computing  has now advanced into the NISQ era~\cite{preskill2018}. Quantum circuits with more than 50 qubits were successfully manipulated to demonstrate the so-called quantum supremacy ~\cite{arute2019,zhong2020,wu2021}.
However, the current state-of-the-art quantum processors are still not able to carry out universal quantum computing.
Particularly, limited lifetime of coherent qubits, environmental noises and connectivity of NISQ devices limit the scale of applications of quantum algorithms. Nevertheless, NISQ devices could be useful for specific tasks.
In recent years, several variational quantum algorithms have been proposed to solve classical or quantum problems by parameterizing quantum circuits using certain classical-quantum hybrid optimization approaches.   This has led to, for example, the successful applications of quantum devices ~\cite{peruzzo2014,grimsley2019,chen2020} in the  classical combinatorial optimization~\cite{farhi2014}, quantum chemistry~\cite{peruzzo2014,colless2018,cao2019,grimsley2019,harsha2018,mcardle2020,taube2006}, quantum machine learning~\cite{benedetti2019,kimura2021,liu2019}, and condensed matter physics~\cite{bravo-prieto2020,cade2020,cai2020,reiner2019,yalouz2021,uvarov2020,grant2019,liu2021}.

VQE is one of the earliest variational quantum algorithms that have been proposed \cite{peruzzo2014}. It was first introduced to solve ground states using NISQ  devices~\cite{peruzzo2014,colless2018,cao2019,grimsley2019,harsha2018,mcardle2020,taube2006,benedetti2019,kimura2021,chen2020}.
Several extended schemes of VQE have also been proposed for computing excited states. This includes the orthogonality constrained VQE (OC-VQE)~\cite{higgott2019}, variational quantum deflation (VQD) algorithm~\cite{wen2021,kuroiwa2021}, subspace search VQE (SS-VQE)~\cite{nakanishi2019}, and multistate contracted VQE (MC-VQE)~\cite{parrish2019}.
The first two algorithms determine the excited states recursively. At each step, one excited state is variationally optimized in the basis space which excludes the ground state as well as other excited states determined in the previous steps.
This recursive scheme demands extra computational resource to optimize the trial state so that it is orthogonalized to the previously determined eigenvectors.
The latter two algorithms also require additional quantum operations after the minimization of the loss function. As a penalty of these extra demands, one has to increase the depth of quantum circuits. This, inevitably, will increase the errors in the optimization process. Hence it remains challenging to determine excited states using NISQ computing devices.

In this work, we propose a novel variational quantum algorithm to simultaneously determine many excited states using a purification technique by introducing ancillary qubits to parameterize and measure their wave functions in one quantum circuit. This algorithm optimizes the Hilbert subspace from which the excited states can be accurately determined and greatly simplifies the steps in the preparation of initial states. The circuit contains both the system qubits used for parameterizing the quantum states to be optimized and the ancillary qubits. Starting from a quantum state in which each ancillary qubit is maximally entangled with a system qubit, our algorithm is able to train and measure many orthogonal trial states at the same time. The circuit implements a unitary transformation for the purified wave function. It keeps the orthogonality of the trial states in every step of evolution, same as in the VQE optimization of the ground state. Furthermore, measurements on the final states of the ancillary and system qubits bestow the ability to determine the expectation values of any physical operators. Our scheme is compatible to the existed VQE algorithms and provides new tool-kits to extend the abilities of post processing controls. It shows great potential to practical applications on quantum processors against noisy implementations on NISQ devices.

\section{Variational eigensolver for excited states}
\label{sec:method}

Implementation of VQE on a NISQ device is to use a quantum circuit to parameterize a unitary transformation through quantum gates for a quantum state so that its energy can be minimized. It starts from an initial state prepared according to a variational ansatz that fulfill the variational principle. A loss function is evaluated by repetitively measuring the final state which is a superposition of all qubit states. Similar measurements are performed to determine the derivatives of the loss function with respect to variational parameters governed by the parameter-shift rules~\cite{mitarai2018,schuld2019}. These measured results of the loss function and its derivatives are then used to optimize the variational parameters. Repeating this optimization scheme for sufficiently many times, we will obtain a set of converged and optimized variational parameters from which the ground state energy can be determined.

The above strategy cannot be directly applied to simultaneously find a set of eigenstates, say $K$ lowest energy eigenstates of a Hamiltonian, using just one quantum circuit. This is because a quantum circuit cannot simultaneously optimize two or more orthogonalized trial states. To resolve this problem, we introduce a set of ancillary qubits (ancillas) to convert these $K$ or more orthogonal trial states into a pure quantum state (this procedure is known as ``purification'' in quantum information). The optimization can then be imposed to this purified quantum state just using one quantum circuit. Purification of quantum states by introducing ancillas is in fact a commonly used technique in physics~\cite{verstraete2004}, particularly in the study of thermodynamics of quantum many-body systems~\cite{kleinmann2006,gullans2020}, as well as in quantum computation~\cite{nielsen2010,endo2020}. This technique has also been invoked to investigate dynamic Green's functions through variationally optimized quantum circuits~\cite{endo2020}. The use of this technique plays a crucial role in our algorithm. It, as will be demonstrated later, can significantly reduce the complexity of the quantum circuit, especially the depth of quantum circuit (or the total number of unitary quantum operations that are needed for optimization), and the errors in the final results. 

\begin{figure*}[t]
\centering
\includegraphics[width=0.85\textwidth]{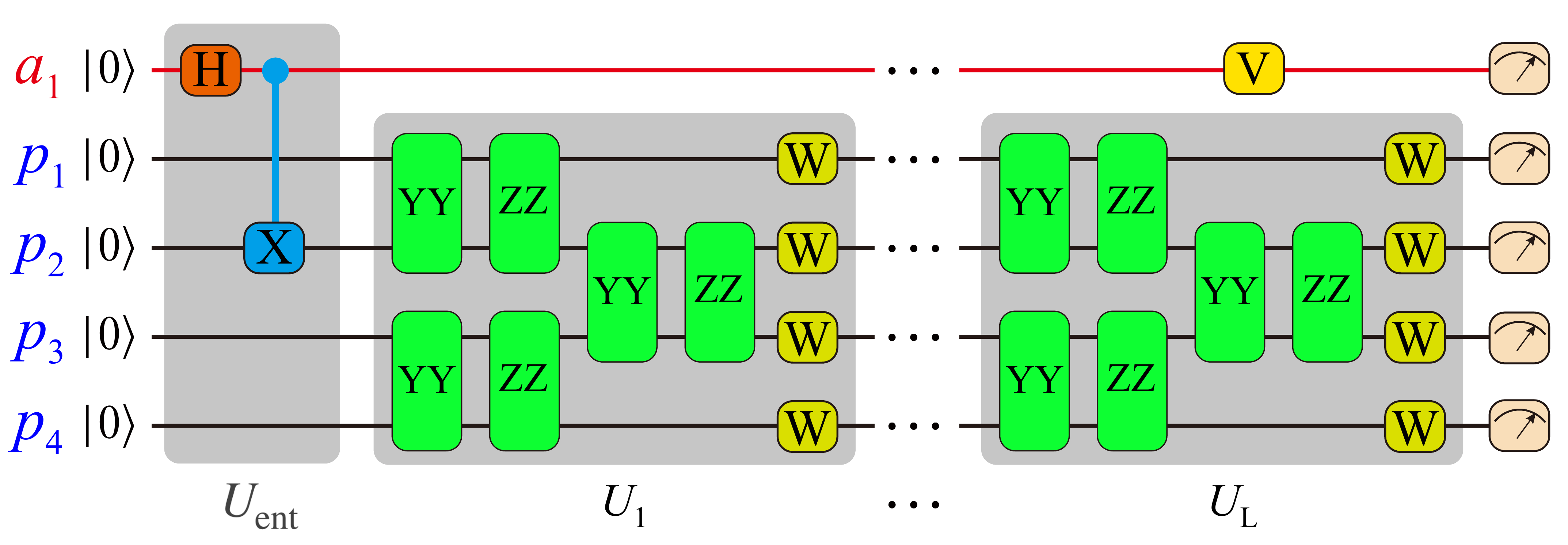} 
\caption{
Sketch of a quantum circuit of variational ansatz for a $N_p=4$ and $N_a=1$ system.
An entanglement generator $U_{\rm ent}$ is applied between the ancillary qubit and the physical qubits to prepare a purified initial state. $U_l$ is a collection of all unitary gate operations, which act only on physical qubits, in the $l$-th layer of the circuit. $V$ denotes a post processing operator, which applies to the ancillary qubit only. $W$ represents a single-qubit gate, which is parameterized by a product of three rotational gates $W(\alpha,\beta,\gamma) = R_X(\alpha) R_Z(\beta) R_X(\gamma)$ in this work. $YY$ and $ZZ$ are the two-qubit gates for implementing unitary rotations along the $y$- and $z$-axis, respectively.
At the end of the circuit, projective measurements are imposed to retrieve the loss function or the expectation values of other physical quantities.}
\label{Fig1}
\end{figure*}

\subsection{The loss function}

Let us consider a system of $N_p$ physical qubits (we call it a physical system) on which the Hamiltonian $\hat H$ is embedded. In order to determine the $K$-lowest energy eigenstates, we first introduce $N_a$-ancillary qubits to construct a purified quantum state from which $K$-orthogonal trial states can be optimized. Here $N_a$ should be larger than or equal to $\log_2 K$ but smaller than $N_p$. The whole system therefore contains $N_p$ physical qubits and $N_a$ ancillas. The variational optimization of this purified quantum state is carried out by performing unitary transformation for the physical qubits only. The ancilla states are not altered by the circuit once they are initialized. Besides the role of purification, another important role that the ancillas play is to ensure the $K$ trial states in the physical system are orthogonal to each other in the whole circuit.

We initialize the system by demanding that the physical qubits form a maximally entangled state with the $N_a$ ancillas
\begin{equation}
\ket{\psi_{\rm ini}} = \frac{1}{\sqrt{M}} \sum_{\alpha=0}^{M-1} \ket{\alpha , p} \otimes \ket{\alpha ,a} , \quad (M  = 2^{N_a}),
\label{eq:init_state}
\end{equation}
where $\{ \ket{\alpha, p} \}$ and $\{ \ket{\alpha,a} \}$ are the $M$ orthogonal basis states of the physical and ancillary systems, respectively. For the ancillary system, the orthogonal basis states can be simply taken as
\begin{equation}
\ket{\alpha,a} = \prod_{i=1}^{N_a} \otimes \ket{\alpha_i,a} , \quad \alpha = (\alpha_1 , \cdots \alpha_{N_a}),
\end{equation}
where $\ket{\alpha_i, a} $ ($\alpha_i = 0$ or $1$) is a basis state of the $i$th ancilla.
This maximally entangle state can be readily prepared by requiring, for example, the first $N_a$ qubits in the physical system each forms a maximally entangled state with a corresponding ancilla, and the rest physical qubits are all in the up-spin states, namely
\begin{equation}
\ket{\psi_{\rm ini}} = \prod_{i=1}^{N_a} \otimes \ket{B}_i  \prod_{j = N_a+1}^{N_p} \otimes \ket{0, p}_j,
\label{Eq:initial}
\end{equation}
where $\ket{B}_i$ is a Bell state formed by the $i$-th physical qubit and the $i$-th ancilla~\cite{nielsen2010}
\begin{equation}
\ket{B}_i = \frac{1}{\sqrt{2}}\left(\ket{0,p} \ket{0,a} + \ket{1,p} \ket{1,a} \right)_i .
\end{equation}

The variational wave function is defined by applying a unitary operation $U(\theta )$ to the initial state
\begin{equation}
\ket{\psi(\theta )} =  U(\theta)\ket{\psi_{\rm ini}} = \frac{1}{\sqrt{M}}\sum_{\alpha} \left[ U(\theta ) \ket{\alpha,p}\right] \ket{\alpha ,a}
\label{Excited_ansatz}
\end{equation}
where $U(\theta )$ acts on the physical qubits only and $\theta$ are  the variational parameters. It is parameterized as a quantum circuit
\begin{equation}
U (\theta ) = \prod_{l=1}^{L} U_{l}(\theta_l) , \quad \theta = (\theta_1, \cdots \theta_p) ,
\end{equation}
where $U_l(\theta _l)$ is a unitary operator that acts on the $l$-th layer of the circuit, and $\theta_l$ is a collection of all variational parameters on that layer.

As an example, Fig.~\ref{Fig1} shows a quantum circuit that realizes a unitary transformation. The circuit consists of both unitary two-qubit gates and single-qubit quantum rotation gates. $\theta$ are essentially the parameters that control these unitary gates. In general, the circuit should be designed or adjusted to make the overlaps between the variational states and the targeted excited states as large as possible, avoiding a barren plateau in the optimization~\cite{mcclean2018}.

The loss function we used is the total energy expectation values of $K$ orthogonal physical states, which can be selected out from the first $K$ ancillary states,  upon  measurements~\cite{nakanishi2019,parrish2019}
\begin{eqnarray}
\mathcal{L}(\theta) &=&   \sum_{\alpha = 0}^{K-1}  \langle \alpha, p| U^{\dagger}(\theta )\hat H  U( \theta ) |\alpha,p\rangle \nonumber \\
&= & \sum_{\alpha=0}^{K-1} \langle \bar\alpha, p|\hat H |\bar\alpha, p \rangle,
\label{LossFunction}
\end{eqnarray}
where
\begin{equation}
|\bar\alpha ,p\rangle =  U( \theta)|\alpha,p\rangle
\end{equation}
is the final states of the physical qubits.
This loss function, according to the generalized Rayleigh-Ritz variational principle~\cite{gross1988}, sets an upper bound on the sum of the $K$-lowest eigenvalues, ($E_0,\cdots, E_{K-1}$), of $\hat H$, i.e.
\begin{equation}
\mathcal{L}(\theta)   \ge \sum_{i=0}^{K-1} E_i . \label{Eq:Ray-Ritz}
\end{equation}
The eigenvalues are assumed to be ascending ordered, $E_0\le E_1 \le \cdots \le E_{K-1}$. $E_0$ is the ground state energy.

To determine the $K$-lowest eigenvalues and the corresponding eigenvectors of $\hat H$, we should minimize the loss function by variationally optimizing all the gate parameters $\theta$. The stability of this optimization scheme is protected by the generalized Rayleigh-Ritz variational principle~\cite{gross1988}.

\subsection{Determination of energy eigenpairs}

After minimizing the loss function, we project the physical qubits onto the subspace spanned by the $K$-lowest energy states of $\hat H$. However, for a given ancillary state $\ket{\alpha,a}$, the corresponding physical state $\ket{\bar\alpha,p}$ generated by the circuit is not automatically an eigenstate of $\hat H$, because the loss function depends on the sum of the $K$-lowest energy states and cannot distinguish any unitary rotation of these states in that basis subspace. Thus to determine the energy eigenstates, we need to determine through measurement not only the diagonal matrix elements of $\hat H$, i.e. $\bra{\bar\alpha,p} \hat H \ket{\bar\alpha,p}$, but also all off-diagonal matrix elements, i.e $\bra{\bar\beta ,p}\hat H \ket{\bar\alpha,p}$ ($\beta \not= \alpha$), in the final basis subspace.

To measure the off-diagonal matrix element of $\hat H$, generally one has to take a unitary transformation to rotate the physical qubits from $\ket{\bar\alpha , p}$ to $\ket{\bar\beta, p}$. In conventional algorithms, it is difficult to perform this transformation~\cite{parrish2019}. For example, to determine all the elements, MC-VQE needs to prepare different ``reference'' states with each reference containing a pair of initial states. The total number of `reference'' states need to be prepared is $2K^2$. In our algorithm, however, as the ancillas are maximally entangled with the physical qubits, we can implement effectively this unitary transformation by rotating the ancillary qubits only. In other words, to change the physical state from $\ket{\bar\alpha, p}$ to $\ket{\bar\beta, p}$, we just need to change the corresponding ancillary state from  $\ket{\alpha, a}$ to $\ket{\beta, a}$. This basis transformation or rotation does not alter the quantum states of the physical qubits. It provides a feasible scheme to retrieve interested physical quantities that are difficult to compute directly.

The basis transformation from $\ket{\alpha ,a}$ to $\ket{\beta, a}$ can be achieved by applying the operator $\ket{\beta, a}\bra{\alpha, a}$ to the ancillary system. For each ancilla, say the $i-$th one, this basis transformation operator takes four possible values, $\ket{\beta_i}\bra{\alpha_i} = (|0\rangle\langle 0| , |0\rangle\langle 1|, |1\rangle\langle 0|, |1\rangle\langle 1|)_i$, depending on the initial and final basis states of the qubit. These four basis transformation operators are related to the four unitary operators, $\hat A^i = (I, \sigma_x, \sigma_y, \sigma_z )_i$, by a unitary transformation
\begin{equation}
 \ket{\beta_i}\bra{\alpha_i} =  \sum_{\mu_i}  v_{\beta_i \alpha_i ,\mu_i } \hat A^i_{\mu_i},
 \label{Eq:unitary}
\end{equation}
where $I$ is the identity matrix, $(\sigma_x, \sigma_y, \sigma_z )$ are the Pauli matrices, and $ v$ is a $4\times 4$ matrix if we regard $(\beta_i \alpha_i)$ as one index
\begin{equation}
 v = \frac{1}{2} \left( \begin{array}{cccc}
 1 & 0 & 0 & 1 \\ 0 & 1 & i & 0 \\ 0 & 1 & -i & 0 \\ 1 & 0 & 0 & -1
 \end{array}\right) .
\end{equation}

It is simple to show that  the expectation value of the operator
$\hat H\otimes \ket{\beta ,a}\bra{\alpha , a}$ in the final state $\ket{\psi (\theta)}$ equals the matrix element of the Hamiltonian
\begin{eqnarray}
H_{\beta , \alpha}&\equiv&\bra{\bar\beta ,p}\hat H \ket{\bar\alpha,p}
 \nonumber \\
&=& M \bra{\psi (\theta) } \left( \hat H \otimes \ket{\beta ,a}\bra{\alpha , a}\right) \ket{\psi (\theta)}
\end{eqnarray}
Substituting Eq. (\ref{Eq:unitary}) into the above equation, we find that
\begin{equation}
H_{\beta , \alpha}=M \sum_{\mu} V_{\beta\alpha , \mu} \bra{\psi (\theta) }  \hat H \otimes \hat A_\mu \ket{\psi (\theta)},
\label{Eq:Hab}
\end{equation}
where
\begin{eqnarray}
\hat A_\mu &=& \prod_{i=1}^{N_a} \otimes \hat A^i_{\mu_i}, \quad \mu = (\mu_1, \cdots , \mu_{N_a}) , \\
V_{\beta\alpha ,\mu} & = & \prod_{i=1}^{N_a} v_{\beta_i\alpha_i ,\mu_i} ,\quad \beta = (\beta_1,\cdots,\beta_{N_a})
\end{eqnarray}

Eq. (\ref{Eq:Hab}) indicates that by measuring the expectation values of the operators $\hat H \otimes \hat A_\mu$ (the total number of these operators is $4^{N_a}$) in the state $\ket{\psi (\theta)}$
\begin{equation}
h_\mu = \bra{\psi (\theta) }  \hat H \otimes \hat A_\mu  \ket{\psi (\theta)} ,
\label{Eq:hmu}
\end{equation}
we can obtain the matrix element of the Hamiltonian
\begin{equation}
H_{\beta , \alpha}= M\sum_\mu V_{\beta\alpha , \mu} h_\mu .
\label{Eq:Hba}
\end{equation}

By diagonalizing $H_{\beta , \alpha}$ with a unitary matrix $S$
\begin{equation}
H_{\beta , \alpha} = \sum_{i=0}^{M-1} S_{\beta ,i} E_i S^*_{\alpha ,i}
\end{equation}
we obtain all the eigenvalues and eigenvectors. If the eigenvalues are ascending ordered, then the first $K$ eigenvalues, $E_i$ ($i = 0, \cdots, K-1$), are what we hope to target. The corresponding eigenstates are given by
\begin{equation}\label{eq:E_trans}
    \ket{E_i,p} = \sum_{\alpha=0}^{M-1} S_{\alpha ,i}\ket{\bar\alpha,p} ,
\end{equation}

Clearly, the above scheme allows us to manipulate the physical eigenstates or their superpositions by just taking actions on the ancillary qubits.
As an example, let us calculate the thermal average of a physical observable $\hat O$ in the physical subspace spanned by the eigenstates of $\hat H$
\begin{equation}
\langle \hat O\rangle = \text{Tr}(\rho \hat O) , \quad
\rho = Z^{-1} \sum_i e^{-\beta E_i} \ket{E_i,p}\bra{E_i,p},
\end{equation}
where $Z= \sum_i e^{-\beta E_i} $ is an approximate partition function and $\beta$ is the inverse temperature. Again, it is difficult to determine this quantity just by measuring the states of the physical qubits. To resolve this problem, let us introduce a diagonal Hermitian operator in the ancillary system
\begin{equation}
\hat T = Z^{-1} \sum_i  e^{-\beta E_i} \ket{E_i,a}\bra{E_i,a} ,
\end{equation}
where
\begin{equation}
    \ket{E_i,a} = \sum_{\alpha=0}^{M-1} S_{\alpha ,i}\ket{\alpha,a} . 
\end{equation}
It is straightforward to show that
\begin{equation}
\label{eq:thermal}
   \langle \hat O\rangle = M \bra{\psi (\theta)} \hat O \otimes  \hat T \ket{\psi(\theta)} .
\end{equation}
Hence the thermal average of $\hat O$ can be obtained just by measuring the expectation value of $\hat O \otimes  \hat T $ on the final state.

\section{Results of simulations}
\label{sec:examples}

\begin{figure}[!htp]
\centering
\includegraphics[width=0.4\textwidth]{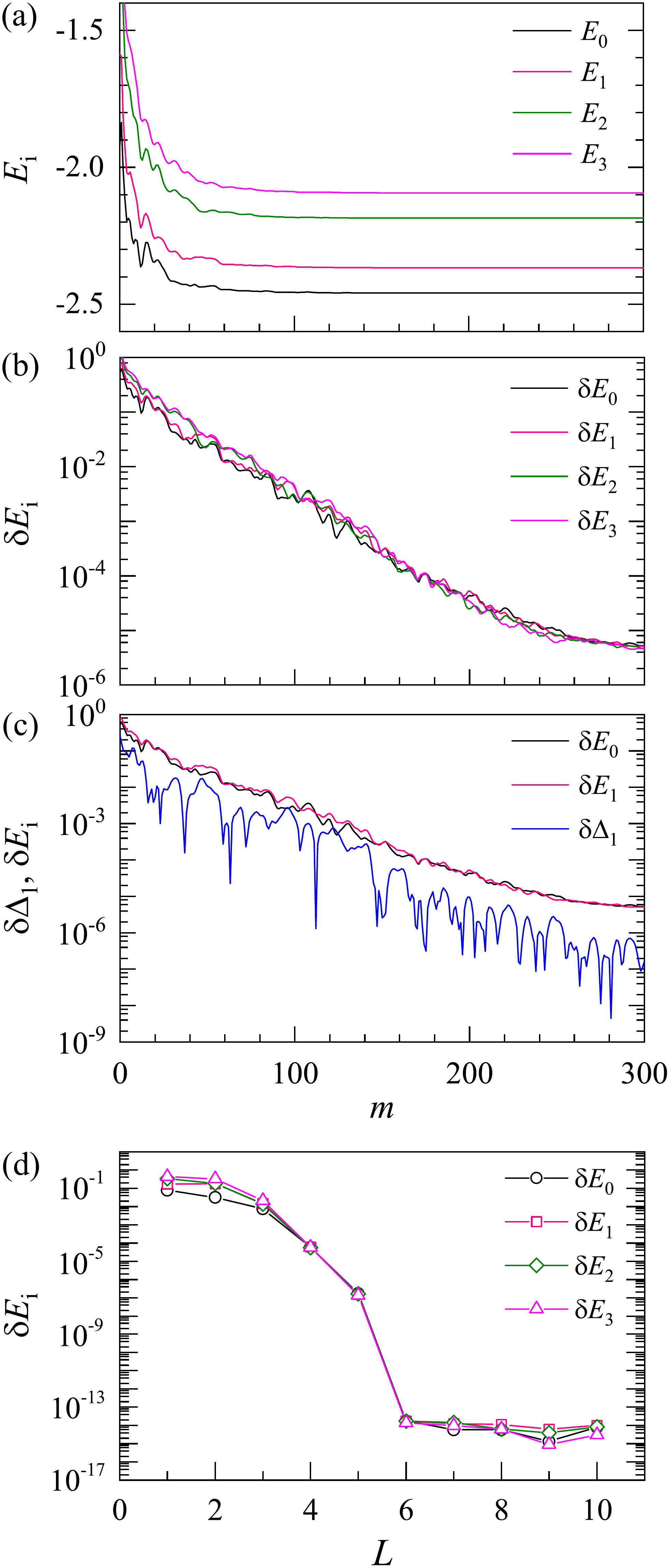} 
\caption{Four lowest eigenvalues obtained with a purified trial wave function using two ancillary qubits of the 8-spin transverse Ising model with $h_x=0.5$ and $J=1$.
(a) and (b) show how the four eigenvalues, $E_i$, and their absolute errors, $\delta E_i = E_i - E_i^\text{ex}$, converge with the iteration number of optimizations $m$, respectively.
Six layers ($L=6$) are used in obtaining the variational results. $E_i^\text{ex}$ is the results obtained by the exact diagonalization.
(c) Comparison of the error of the first energy excitation gap $\delta \Delta_{1} = |\delta E_1-\delta E_0|$ with the errors of the two eigen-energies. Similar results are found for the other excited eigenstates.
(d) Layer number $L$ dependence of the absolute errors of the four converged eigenvalues. In obtaining the converged eigenvalues, $m=6001$ iterations are used.
}
\label{fig:N8A2}
\end{figure}

As an application and demonstration, we apply our method to the one-dimensional quantum transverse Ising model
\begin{equation}
\label{eq:H}
H = - J \sum_i S_{i}^z  S_{i+1}^z + h_x \sum_i S_i^x .
\end{equation}
$S_i^x$ and $S_i^z$ are the $S=1/2$ spin operators. This model is exactly soluble. It allows to compare the results obtained by simulation using our algorithm with the exact ones. We do the calculation at the point, $h_x=0.5$ and $J=1$, where the ground state of this model becomes critical in the thermodynamic limit.

Figure~\ref{Fig1} shows an example of the initial state prepared for a system of $N_a = 1$ and $N_p=4$. In this case, the ancillary qubit is maximally entangled with the physical qubit labelled by $p_2$ in the initial state. The variational ansatz at one layer used in our calculation is~\cite{wecker2015,reiner2019},
\begin{equation}
\label{eq:ansatz}
    U_l (\theta_l) =  W(\theta_{l})  R_{ZZ} (\theta_{l,2}) R_{YY} (\theta_{l,1})
\end{equation}
where $W$ represents a product of three single-qubit rotation gates,
\begin{equation}
    W(\theta_{l}) =  R_X (\theta_{l,5}) R_Z (\theta_{l,4}) R_X (\theta_{l,3}) ,
\end{equation}
and $(X,Y,Z)$ represent the three Pauli matrices $(\sigma_x, \sigma_y,\sigma_z)$.
$R_Q$ is a unitary gate
\begin{equation}
R_{Q} (\theta) = e^{-i \theta Q/2} .
\end{equation}
Each layer comprises two sublayers of two-qubit gates, $R_{YY}$ and $R_{ ZZ}$, in a brick-wall structure, and three sublayers of single qubit gates, $R_{X}$, $R_{Z}$ and $R_{X}$. Fig.~\ref{Fig1} shows the first layer $U_1$ and the last layer $U_L$. Each unitary gate is parameterized by one variational parameter.


Let us first solve 4 lowest eigenstates in a 8-spin system. The dimension of the full Hamiltonian is 256. We use two ancillas to purify the quantum basis states. The variational parameters $\theta$ are randomly initialized within an interval $[0, 0.1)$. The optimization is iteratively conducted for a few hundred times until all the variational parameters are converged. We run the full cycles of optimizations for 21 times, starting from different initial variational parameters, and use the best set of parameters that minimizes the loss function to evaluate  the four lowest eigenpairs of the Hamiltonian.

To determine the matrix elements of the Hamiltonian, we first evaluate the expectation values ($h_\mu$) of the $4^2=16$ operators $\hat H\otimes \hat A_\mu$ defined in Eq. (\ref{Eq:hmu}). Substituting the results such obtained into Eq. (\ref{Eq:Hba}), we then obtain all the $4\times 4$ matrix elements of $\hat H$ in the optimized basis subspace. The four lowest eigenpairs are obtained by diagonalizing this $4\times 4$ matrix.

Fig.~\ref{fig:N8A2}(a) shows how the four eigenvalues, $E_i$ ($i=0,\cdots ,3$), vary with the optimization step $m$ for the transverse Ising model. The four eigenvalues converge very quickly with the increase of $m$. Particularly, as shown in Fig.~\ref{fig:N8A2}(b), their errors with respect to the eigenvalues $E_i^\text{ex}$ obtained by the exact diagonalization, $\delta E_i = E_i - E_i^\text{ex}$, drop exponentially and at nearly the same speed with $m$. Hence the four eigenstates such obtained have the same order of accuracy. It suggests that the energy difference between any two eigenvalues are more accurately calculated than the eigenvalues themselves, which is indeed what we see (Fig. \ref{fig:N8A2}(c)).

Fig.~\ref{fig:N8A2}(d) shows how the errors of eigenvalues, obtained with the best set of converged variational parameters at $m=6001$, vary with the total number of layers $L$ of the circuit. When $L$ becomes larger than 5, the errors of the eigenvalues are determined just by the rounding errors and cannot be further reduced by simply increasing the value of $L$.

\begin{figure}[tp]
\centering
\includegraphics[width = 0.4\textwidth]{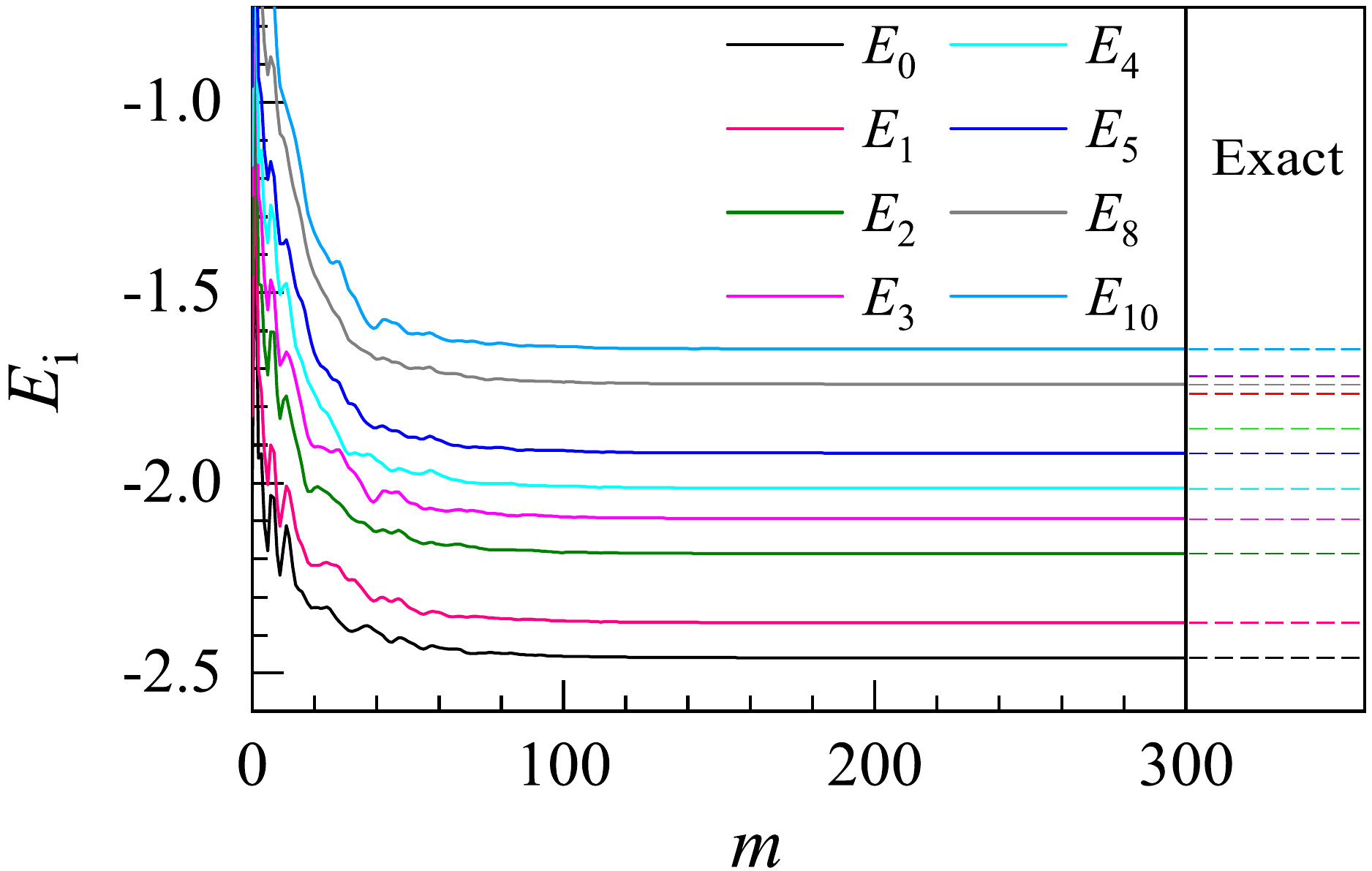} 
\caption{
Eight low-energy eigenvalues obtained using 3 ancillary qubits for the 8-spin transverse Ising model.
The left panel shows how the eight eigenvalues vary with the optimization step $m$.
The right panel shows the results for the lowest 11 eigenvalues of the model obtained by the exact diagonalization. Among these 11 eigenvalues, the seventh, eighth, and tenth eigenvalues are skipped in the variational calculation. }
\label{fig:N8A3}
\end{figure}

\begin{figure}[htp]
\centering
\includegraphics[width = 0.4\textwidth]{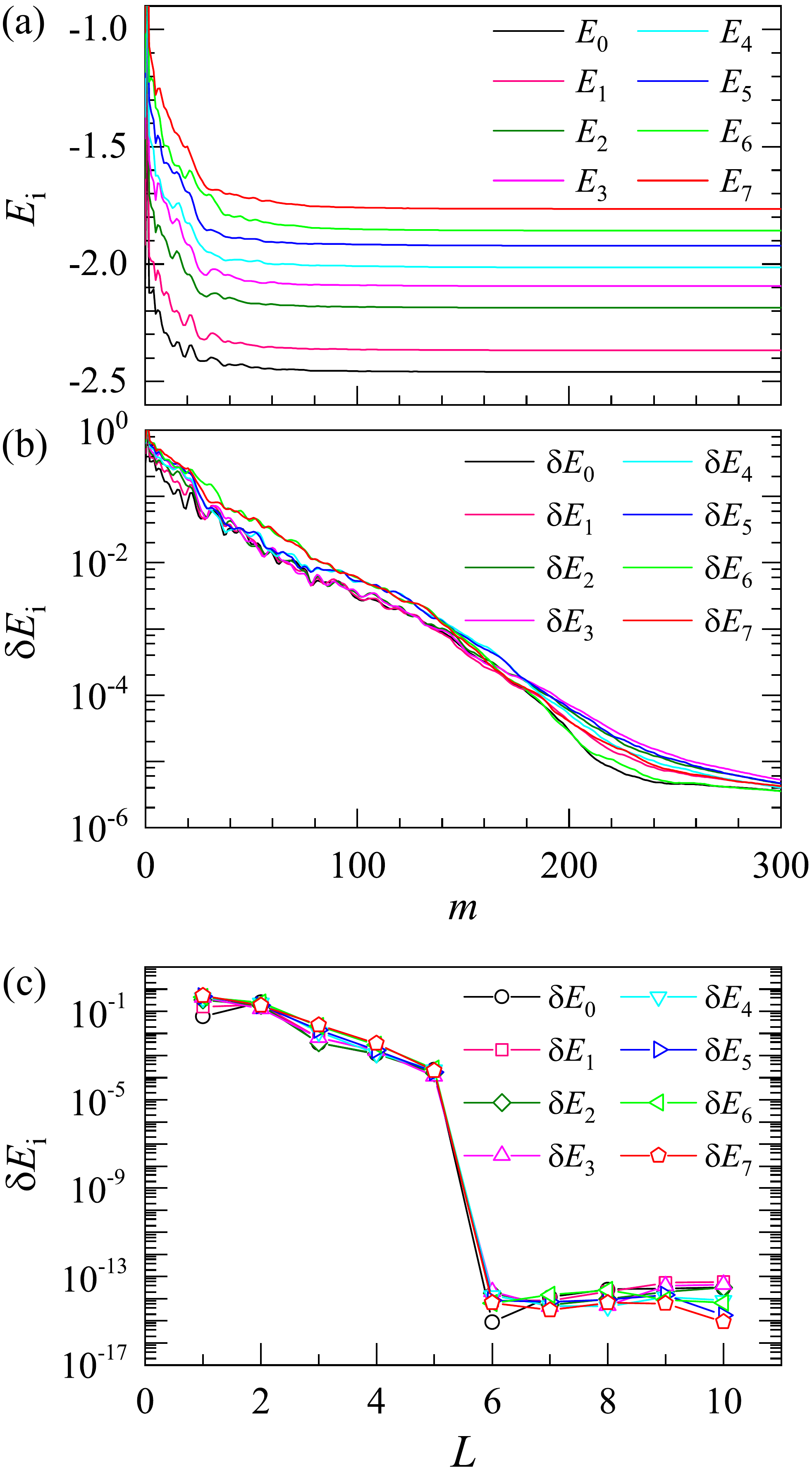} 
\caption{Eight lowest eigenvectors of the transverse Ising model obtained by the variational calculation using 4 ancillary qubits.
(a) and (b)  show how the eight eigenvalues, $E_i$ and their absolute errors, $\delta E_i= E_i-E_i^\text{ex}$, converge with the optimization step $m$. Six variational layers ($L=6$) are used. $E_i^\text{ex}$ is the exact diagonalization result. (c) shows how the absolute errors of the converged eigenvalues, obtained by taking $m=6001$ steps, of optimizations change with the variational layer number $L$.}
\label{fig:N8A4}
\end{figure}

Now let us increase the number of eigenstates to be variationally optimized to eight and see how the algorithm works. We first initialize the wave function using the variational ansatz given in Eq. (\ref{Eq:initial}) with three ancillas. Again, the variational parameters are iteratively optimized by minimizing the loss function.
After evaluating and diagonalizing the Hamiltonian matrix using the optimized wave function, we obtain eight eigenpairs of the Hamiltonian in the subspace spanned by the initial eight basis vectors.

Figure~\ref{fig:N8A3} shows how the eight eigenvalues converge with the iteration number $m$. The dashed horizontal lines on the right panel of the figure are the eigen-energies of the Hamiltonian obtained by the exact diagonalization. Among the eight eigenvalues we obtain, six of them, including the ground state energy and the energies of the lowest five excited states, converge to the exact results. The other two eigenvalues, however, converge to the eighth  and tenth excited eigenvalues of the full Hamiltonian. Hence the sixth and seventh excited eigenstates are missing in this calculation.

To understand why the variational wave function optimized through the quantum circuit does not produce the eight lowest energy eigenstates, we evaluate the wave function overlaps between the eight lowest energy eigenstates of the Hamiltonian and the eight initial physical basis states  $\ket{\alpha , p}$ ($\alpha = 0, \cdots , 7$)
\begin{equation}
    M_{i, \alpha} = \braket{E_i^\text{ex} | \alpha, p} .
\end{equation}
In order to obtain all eight lowest energy eigenstates of $\hat H$, $M$ should be a matrix of rank 8. However, in our calculation, we find that its rank is 6. It indicates that only 6 out of the 8 lowest energy eigenstates of $\hat H$ can be obtained from the variational ansatz we adopt if only three ancillary qubits are used.

If the variational ansatz is not changed in our layered circuit structures, one way to solve the above problem is to increase the number of ancillas. By introducing more ancillary qubits, we are able to optimize the wave function in a  larger Hilbert subspace. By increasing the number of ancillas, we should be able to find the eight lowest eigenpairs. In other words, if the eigenpairs such obtained do not change with the increase of the number of ancillary qubits, they should be the targeted eight lowest eigenpairs.

For the eight-spin transverse Ising model, we find that it is sufficient to find the eight lowest energy eigenstates by utilizing 4 ancillas.
Fig.~\ref{fig:N8A4}(a) shows how the lowest eight eigenvalues of $\hat H$ converge with the iteration step $m$. The absolute errors $\delta E_i$, ($i=0,\cdots,  7$) of these eigenvalues, shown in Fig.~\ref{fig:N8A4}(b), drop exponentially and at almost the same speed with $m$. For all the eight eigenvalues, the errors become less than $10^{-6}$ if six layers ($L=6$) are used and $m$ becomes larger than 300. The errors can be further reduced if more iteration steps are taken to optimize the variational parameters.   Fig.~\ref{fig:N8A4}(c) shows the $L$-dependence of the converged error $\delta E_i$ obtained by taking $m=6001$ steps of optimizations.

\section{Discussion and Summary}
\label{sec:summary}

We have proposed a novel variational quantum algorithm for computing low-energy excited states of a Hamiltonian using a quantum circuit. This algorithm lowers dramatically the steps in the preparation of initial trial wave functions, the depth of the circuit for optimizations, and the complexity in the measurement of physical observables by utilizing the purification technique. The targeted excited states are purified by coupling with several ancillary qubits to form a maximally entangled state so that only one initial trial state, such as the state defined by Eq. (\ref{Eq:initial}), needs to be prepared.
Moreover, only one loss function, constructed based on the generalized Rayleigh-Ritz variational principle~\cite{gross1988},  needs to be minimized to train all the variational parameters no matter how many excited states are computed. In other words, our algorithm has the same computational complexity as the standard VQE used in the variational optimization of the ground state only.
Furthermore, as our algorithm allows a quantum circuit with a relatively shallow depth to be used to achieve the same accuracy as other VQE algorithms previously introduced~\cite{higgott2019,wen2021,kuroiwa2021,nakanishi2019,parrish2019}, it reduces the total errors in the gate operations as well as the readout errors, and promises to be resilient of noisy implementations on NISQ hardwares.

In the implementation of the algorithm, the variational unitary gate optimization is deployed on the physical qubits only. As a unitary transformation does not alter the maximally entangled nature between the physical and ancillary qubits, the orthogonality of the targeted excited states is preserved throughout the optimization. This orthogonality preserving feature could be used to improve the trainablity, efficiency and accuracy of variational quantum eigensolvers for computing excited states with a quantum circuit.

To variationally determine the $n$-lowest energy eigenstates, the initial physical basis states, which are imbedded in the maximally entangled wave function, should have finite overlaps with the $n$-lowest eigenstates of $\hat H$. Otherwise, some eigenstates may converge to higher energy states of $\hat H$. This problem can be resolved simply by increasing the number of ancillas so that a broader excited-state spectrum can be covered. One can also exam whether the states determined from the variational optimization are truly the lowest energy eigenstates by increasing the number of ancillas. If all the eigenpairs computed do not change with the increase of the number of ancillas, these states are most likely to be the targeted lowest energy eigenstates within the optimization and measurement errors.

The maximally entangled feature of the optimized quantum state grants access to any excited states or their superpositions in the targeted physical subspace just by rotating ancillary qubits. It extends significantly the capability of the VQE methods and allows us to determine all the matrix elements (including the off-diagonal ones) of the Hamiltonian or other physical observables, without increasing the depth of quantum circuit. This is particularly useful in the practical application of NISQ computing devices.

The results we obtained show unambiguously that the eigenvalues obtained with our algorithm converge unanimously with the optimization steps and have the same order of accuracy. As a consequence of this, the energy difference between any two eigenvalues obtained with this algorithm has a higher accuracy than the eigenvalues themselves. This is a feature that is not observed in other VQE algorithms for calculating low energy excited states of a Hamiltonian. The constrained VQE methods \cite{higgott2019,wen2021,kuroiwa2021}, for example, suffers the problem of cumulative errors, which implies that higher excited states obtained with these methods have higher errors. Several algorithms \cite{nakanishi2019,parrish2019} were proposed to resolve this problem by searching the low energy eigenspace before evaluating eigenstates. However, to generate the target eigenstates, these approaches needs to deploy a second stage of optimization with more variational parameters after the first one. The second stage of optimization will surely increase the circuit depth and lower the accuracy of the results.

The implementation of our algorithm does not require  all-to-all qubit connectivity of quantum hardware which is still a challenge for NISQ devices. 
For instance, the algorithm can be readily implemented on a NISQ device with a  ladder-shaped architecture which now could be realized as a subset of square lattice quantum devices~\cite{arute2019,zhong2020,wu2021}. 
Qubits on one of the ladders are considered as the physical qubits and those on the other leg are the ancillas. 
With the rapid development of NISQ devices in connectivity as well as in reliability, our algorithm can be implemented with high fidelity.
Having the noise resilient features and the feasible post processing tool-kits, our algorithm holds great promise for eigenpairs determinations and various applications on upcoming NISQ devices and fault-tolerant quantum computers.

\vspace{5mm}
\noindent {\bf Acknowledgements:}

\noindent This work is supported by the National Key Research and Development Project of China (Grant No.~2017YFA0302901),
the National Natural Science Foundation of China (Grants No.~11888101 and No.~11874095),
the Youth Innovation Promotion Association CAS (Grants No.~2021004), and the Strategic
Priority Research Program of Chinese Academy of Sciences (Grant No.~XDB33010100 and No.~XDB33020300).

\bibliographystyle{apsrev4-2}
\bibliography{avqe}

\end{document}